\documentclass[reprint,showpacs,
superscriptaddress,
 amsmath,amssymb,
 aps,prl
]{revtex4-1}

\usepackage{graphicx} 
\usepackage{dcolumn}
\usepackage{bm}

\begin{document}

\title{Dynamics and symmetries of flow reversals in turbulent convection}

\author{Mani chandra}
\affiliation{Department of Physics, Indian Institute of Technology, Kanpur, India 208016}
\author{Mahendra K. Verma}
\email{mkv@iitk.ac.in}
\affiliation{Department of Physics, Indian Institute of Technology, Kanpur, India 208016}
\date{\today}

\begin{abstract}
Based on direct numerical simulations and symmetry arguments, we show that the large-scale Fourier modes are useful tools to describe the flow structures and dynamics of flow reversals in Rayleigh-B\'enard convection (RBC).     We observe that during the reversals, the amplitude of one of the large-scale modes vanishes, while another mode rises sharply, very similar to the ``cessation-led'' reversals observed earlier in experiments and numerical simulations.    We find anomalous fluctuations in the Nusselt number during the reversals. Using the structures of the RBC equations in the Fourier space, we deduce two symmetry transformations that leave the equations invariant.  These symmetry transformations help us in identifying the reversing and non-reversing Fourier modes.
\end{abstract}

\maketitle

Many experiments~\cite{Sugiyama:PRL2010,Cioni:JFM1997, Niemela:JFM2001, Brown:JFM2006, Xi:PRE2007, Yanagisawa:PRE2010, Gallet:Arxiv2011} and numerical simulations~
\cite{Sugiyama:PRL2010,Benzi:EPL2008, Breuer:EPL2009,Mishra:JFM2011} on turbulent convection reveal that the velocity field of the system reverses randomly in time (also see review articles~\cite{Ahlers:RMP2009}). This phenomenon, known as ``flow reversal", remains ill understood.   This process gains practical importance due to its similarities with the magnetic field reversals in geodynamo  and  solar dynamo~\cite{Glatzmaier:NATURE1995}.   In this letter, we study the dynamics and symmetries of flow reversals in turbulent convection using the large-scale Fourier modes of the velocity and temperature fields.

The experiments and simulations performed to explore the nature of flow reversals are typically for an idealized convective system called Rayleigh-B\'{e}nard convection (RBC) in which a fluid confined between two plates is heated from below and cooled at the top.    Detailed measurements show that the first Fourier mode vanishes abruptly during some reversals~\cite{Brown:JFM2006,Xi:PRE2007}.  These reversals are referred to as ``cessation-led".  Recently Sugiyama {\em et al.}~\cite{Sugiyama:PRL2010} performed RBC experiments on water in a quasi two-dimensional box, and observed flow reversals with the flow profile dominated by a diagonal large-scale roll and two smaller secondary rolls at the corners.  They attribute the flow reversals to the growth of the two smaller corner rolls as a result of plume detachments from the boundary layers.  

Several theoretical studies performed to understand reversals in RBC provide important clues.  Broadly, these works involve either stochasticity (e.g., ``stochastic resonance"~\cite{Sreenivasan:PRE2002,Benzi:EPL2008}), or low-dimensional models with noise~\cite{Araujo:PRL2005,Brown:PF2008}.   Mishra {\em et al.}~\cite{Mishra:JFM2011} studied the large-scale modes of RBC  in a cylindrical geometry 
and showed that the dipolar mode decreases in amplitude and the quadrupolar mode increases during the cessation-led reversals. Regarding dynamo, low-dimensional models constructed using the large-scale modes and symmetry arguments reproduced dynamo reversals successfully~\cite{Petrelis:PRL2009,Gallet:Arxiv2011}.     

The theoretical models described above only focus on the large-scale modes.  Here too, they provide limited information about these modes due to small number of measuring probes.   In this letter we compute large-scale and intermediate-scale Fourier modes accurately using the complete flow profile. This helps us in quantitative understanding of the dynamics and symmetries of the RBC system. We also show that these modes can describe the diagonal and corner rolls of Sugiyama {\em et al.}~\cite{Sugiyama:PRL2010}, as well as the cessation-led reversals of Brown and Ahlers~\cite{Brown:JFM2006}.  
The properties of the modes for convection are contrasted with those for dynamo.

The equations governing two-dimensional Rayleigh-B\'enard convection under Boussinesq approximation are
\begin{eqnarray}
    \frac{\partial \mathbf{u} }{\partial t} + (\mathbf{u} \cdot \nabla)
    \mathbf{u} & = & -\nabla P + Pr \nabla^{2} \mathbf{u} + Ra Pr T
    \hat{y},  \label{eqn:vel_eqn} \\ 
    \frac{\partial T}{\partial t} +
    (\mathbf{u} \cdot \nabla) T & = & \nabla^{2} T, \\ \label{eqn:temp_eqn}
    \nabla \cdot \mathbf{u} & = & 0, \label{eqn:div_free_condition}
\end{eqnarray}
where $\mathbf{u}$ is the velocity field, $T$ is the temperature field, $P$ is the pressure, and $\hat{y}$ is the buoyancy direction. The two nondimensional parameters are the Prandtl number $Pr$, the ratio of the kinematic viscosity $\nu$ and the thermal diffusivity $\kappa$, and the Rayleigh number $Ra = \alpha g \Delta d^3/(\nu \kappa)$, where $\alpha$ is the thermal expansion coefficient, $d$ is the distance between the two plates, $\Delta$ is the temperature difference between the plates, and $g$ is the acceleration due to gravity.  The above equations have been nondimensionalized using $d$ as the length scale, the thermal diffusive time $d^2/\kappa$ as the time scale, and $\Delta$ as the temperature scale.

\begin{figure*}[htbp]
    \begin{center}
        \includegraphics[scale=0.3]{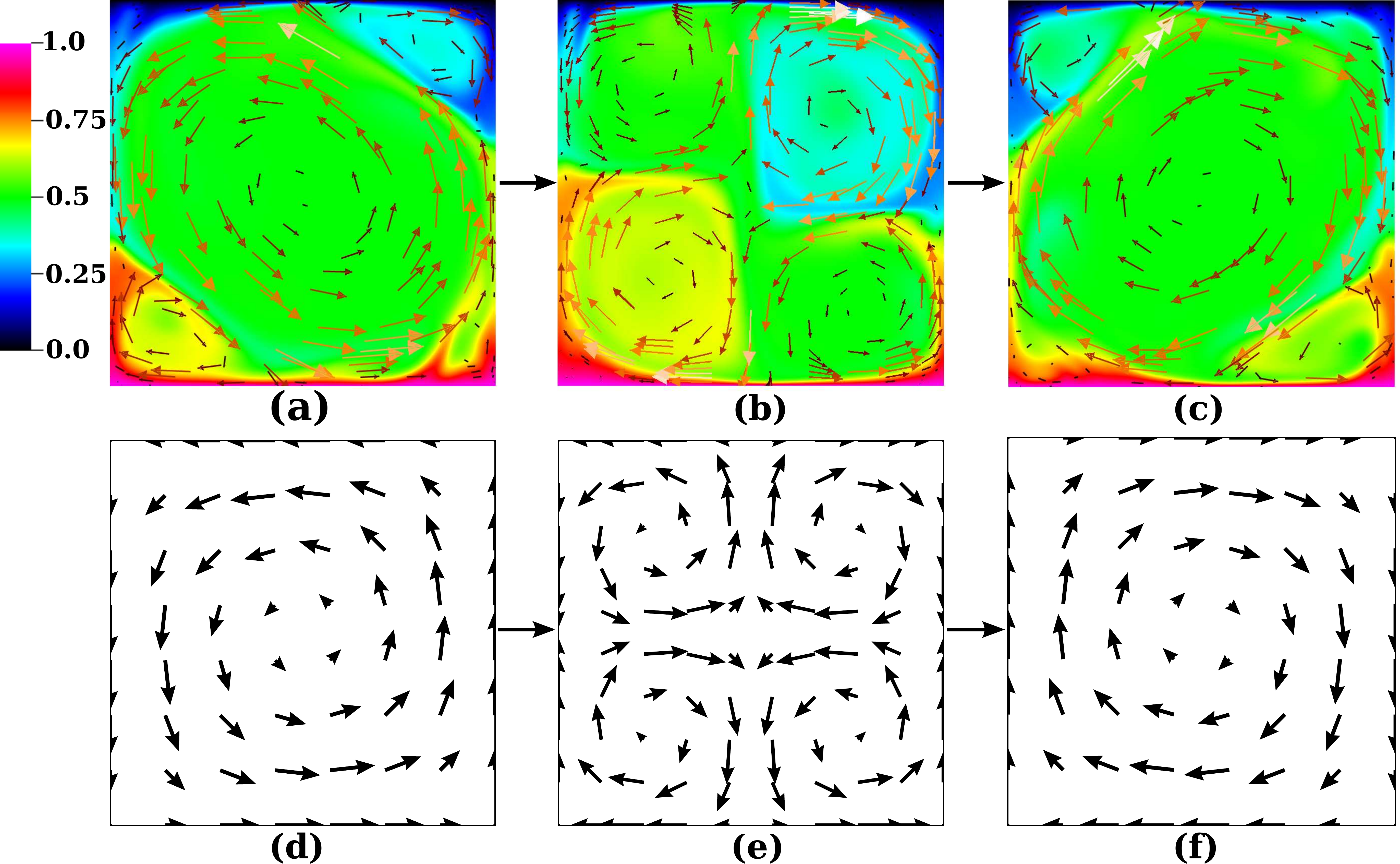}
    \end{center}
    \caption{Top panel: Convective flow profile computed from a spectral element simulation for a box of aspect ratio $\Gamma = 1$, Prandtl number $Pr = 1$, Rayleigh number $Ra = 2 \times 10^7$: (a) before the reversal ($t=10.52$ thermal diffusive time units), (b) during the reversal ($t=10.68$), (c) after the reversal ($t=10.79$).  Bottom panel:  Velocity field corresponding to (a) ${\bf \hat{u}}_{1,1}$,  (b)${\bf \hat{u}}_{2,2}$, (c)  $-{\bf \hat{u}}_{1,1}$.  The flow patterns of the bottom panel are good approximations of the corresponding patterns of the top panel.  The corner rolls of (a,c) are obtained as a result of superposition of (d,f) with (e).   }
    \label{fig:AR_1_real_space_fields}
\end{figure*}

We solve the above equations in a closed box geometry of aspect ratio $\Gamma =1$ (denoted by $B_1$) and $\Gamma = 2$ (denoted by $B_2$) using Nek5000~\cite{Fischer:JCP1997}, an open source  spectral-element code.   We apply no-slip boundary condition on all the walls.  The top and bottom walls are assumed to be perfectly conducting, while the side walls are assumed to be insulating.     We use $28\times28$ spectral elements for $B_1$ and $48\times28$ spectral elements for $B_2$, and $7^{th}$ order polynomials for resolution inside the elements.  Thus, the effective grid resolution for the $B_1$ and $B_2$ boxes are $196\times 196$  and $336\times 196$ respectively.  The concentration of grid points is higher near the boundaries in order to resolve the boundary layer.   We perform our simulations for $Ra = 2 \times10^7$, $10^8$, $10^9$ for $B_1$, and for $Ra$ = $10^7$, $2 \times 10^7$, $10^8$ for $B_2$, till several thermal diffusive time units.  $Pr=1$ for all our runs. We observe flow reversals only for $Ra = 2 \times10^7 (B_1)$ and $Ra = 10^7 (B_2)$.  These results are  consistent with those of Sugiyama {\em et al.}~\cite{Sugiyama:PRL2010}.

 In Fig.~\ref{fig:AR_1_real_space_fields}(a,b,c) we display three frames of the velocity and temperature fields for the $Ra = 2 \times10^7$ run for the $\Gamma=1$ box.  The three frames illustrate the flow profiles before the reversal ((a), at $t=10.52$ thermal diffusive time units), during the reversal ((b), at $t=10.68$), and after the reversal ((c), at $t=10.79$) (see videos in~\cite{movie}).  They are very similar to those presented by Sugiyama {\em et al.}~\cite{Sugiyama:PRL2010}.   To gain further insights into the reversal dynamics, we decompose the velocity and temperature fields into Fourier modes 
\begin{eqnarray}
    u & = & \sum_{m,n} \hat{u}_{m,n} \sin(m k_c x) \cos(n \pi y), \\
    v & = & \sum_{m,n} \hat{v}_{m,n} \cos(m k_c x) \sin(n \pi y), \\
    T & = & \sum_{m,n} \hat{T}_{m,n} \cos(m k_c x) \sin(n \pi y),
\end{eqnarray}
where ${\bf u} = (u,v)$ is the velocity field, $k_c$ = $\pi$ for $\Gamma=1$, and $k_c$ = $\pi/2$ for $\Gamma=2$.    

We compute the Fourier amplitudes using FFTW library through  PyFFTW interface~\cite{fftw}.  For these transforms we interpolate the Nek5000 data to a uniform $128 \times 128$ grid. In Fig.~\ref{fig:AR_1_real_space_fields}(d,e) we display the velocity profiles of the primary modes ${\bf k} = (1,1)$  and $(2,2)$, which are a single roll, and four rolls respectively.  The sign of the Fourier mode of Fig.~\ref{fig:AR_1_real_space_fields}(f) is reversed compared to that of Fig.~\ref{fig:AR_1_real_space_fields}(d).  The diagonally oriented roll and the corner rolls of Fig.~\ref{fig:AR_1_real_space_fields}(a,c) can be well approximated as a superposition of Fig.~\ref{fig:AR_1_real_space_fields}(d,f)  and Fig.~\ref{fig:AR_1_real_space_fields}(e)  with appropriate amplitudes.  The corresponding primary modes for $\Gamma=2$ box are ${\bf k} = (2,1)$ and $(2,2)$.

\begin{figure}
   \begin{center}
        \includegraphics[width=8.6cm, height=!]{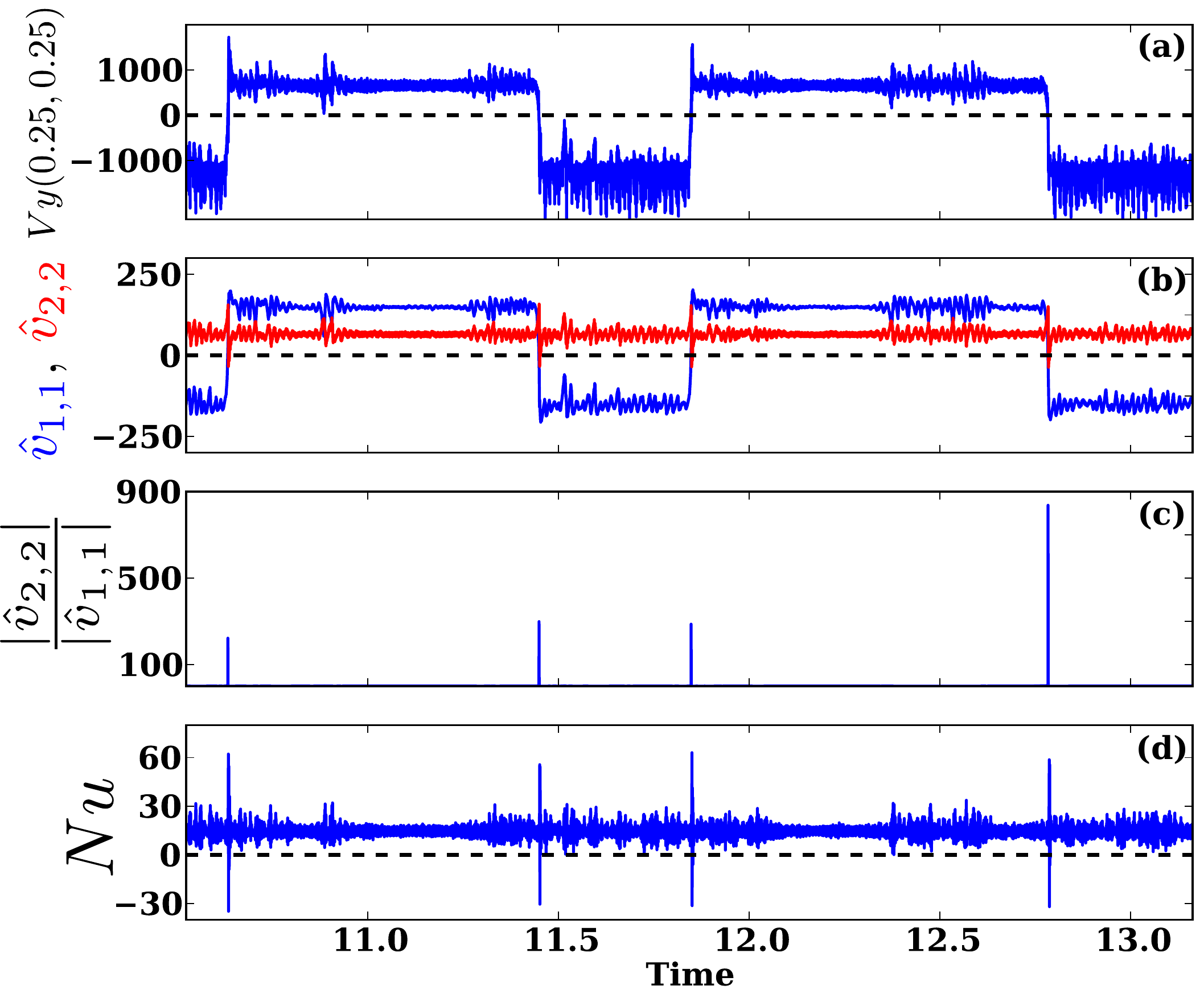}
    \end{center}
    \caption{For $\Gamma =1$, $Pr = 1$, and $Ra = 2\times 10^7$,  time series of (a) $y$ component of the real velocity field at the point $(0.25,0.25)$,  (b) the Fourier amplitudes $ \hat{v}_{1,1}$ (blue) and $ \hat{v}_{2,2} $ (red), (c) the ratio $|\hat{v}_{2,2}/\hat{v}_{1,1}|$, and (d) Nusselt number ($Nu$).  During the reversal, the sign of mode $\hat{v}_{1,1}$ changes sign after vanishing, while the mode $\hat{v}_{2,2}$ rises sharply.   The Nusselt number shows large fluctuations during the reversals. }
    \label{fig:AR_1_timeseries}
\end{figure}

In Fig.~\ref{fig:AR_1_timeseries} we plot time series of (a) the vertical velocity $V_y$ at $(0.25,0.25)$,  (b) the dominant Fourier modes $\hat{v}_{1,1}, \hat{v}_{2,2}$, (c) the ratio $|\hat{v}_{2,2}/\hat{v}_{1,1}|$, and (d) the Nusselt number, for $\Gamma=1$.    The time series indicates that the mode ${\bf k} = (1,1)$ dominates   the mode $(2,2)$ between the reversals.  During the reversals however, the mode $\hat{v}_{1,1}$ crosses zero, while  $\hat{v}_{2,2}$ shows a spike.  The mode $\hat{v}_{1,1}$ overshoots around 40\% before it attains the steady-state.  Vanishing of  $\hat{v}_{1,1}$, and the peaking of  $\hat{v}_{2,2}$  (the corner rolls)  are the reasons for the four rolls appearing during the reversal (Fig.~\ref{fig:AR_1_real_space_fields}(b)).  This phenomenon is the same as the cessation-led flow reversals reported by Brown and Ahlers~\cite{Brown:JFM2006} and  Mishra {\em et al.}~\cite{Mishra:JFM2011}. 

The mode $(1,1)$ changes sign due to the reversals but the mode $(2,2)$ does not. This is due to certain symmetries obeyed by the governing equations which will be discussed later. As a result, the velocity field near the corners does not reverse (see Fig.~\ref{fig:AR_1_real_space_fields}).   Fig.~\ref{fig:AR_1_timeseries} also shows that the Nusselt number becomes negative during the reversals, i.e., the fluid loses heat energy to the plates for a small time interval while the flow reverses.

\begin{figure}[ht]
    \begin{center}
        \includegraphics[width=8.6cm, height=!]{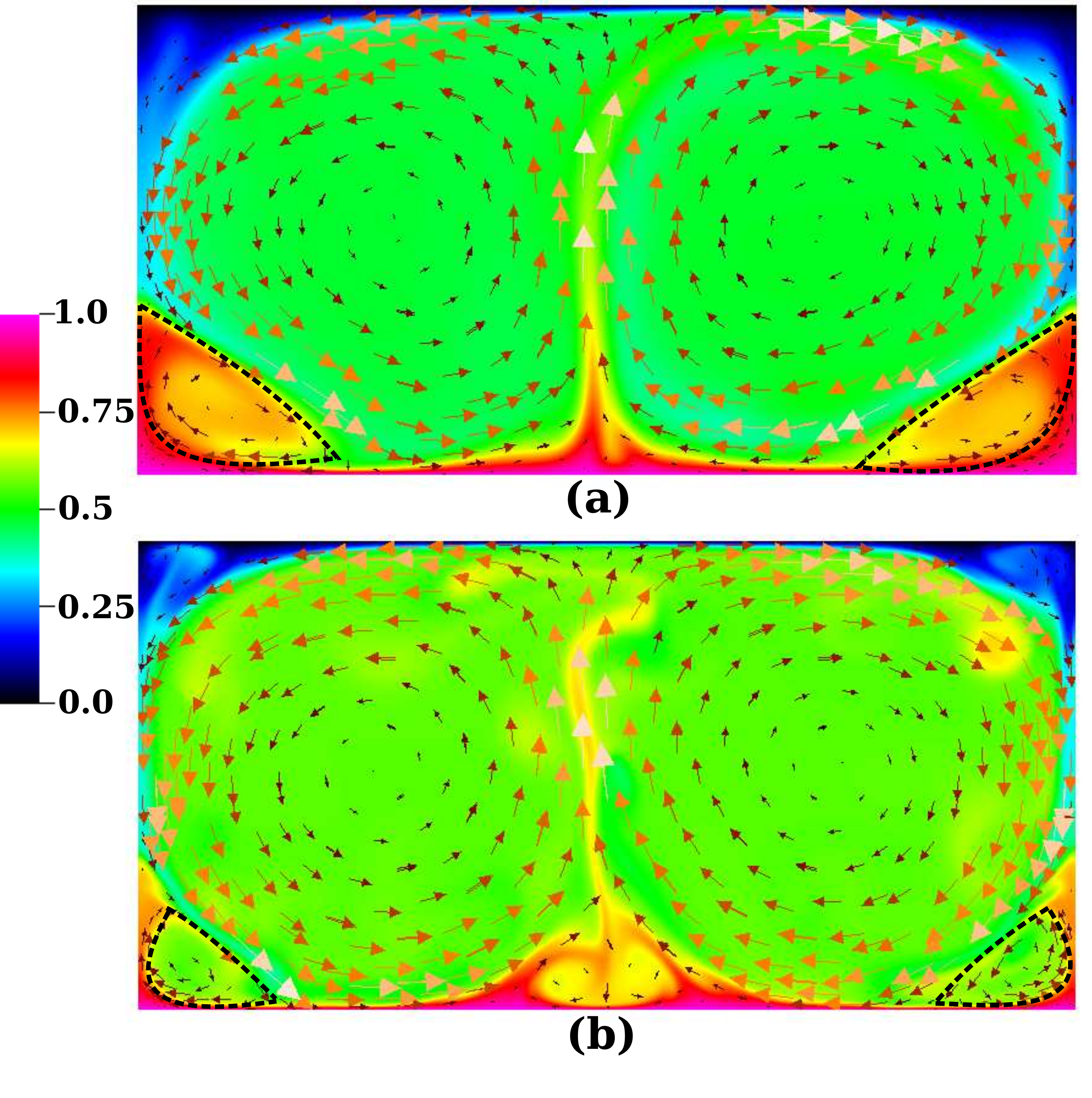}
    \end{center}
    \caption{Convective flow profile for $\Gamma=2$, $Pr=1$,
     (a) reversing case $Ra=10^7$, and (b) non-reversing case $Ra=10^8$.
    The corner rolls of subfigure (a) are much larger than those of subfigure (b). The ratios of the time-averaged Fourier amplitudes $\langle |\hat{v}_{2,2}|  \rangle_t / \langle|\hat{v}_{2,1}| \rangle_t$ for the cases are  0.12(a) and 0.04(b). Small amplitude of $\hat{v}_{2,2}$ makes the growth of the corner rolls difficult.  }
    \label{fig:AR_2_real_space_fields}
\end{figure}

 For the $\Gamma=2$ box, the dynamics of the modes are exactly the same as above, except that the Fourier mode $  (2,1)$ takes the role of the $(1,1)$ mode.  That is, the primary modes for the $\Gamma=2$ box are $(2,1)$ and $ (2,2)$, which represent the two large rolls, and the corner rolls respectively of Fig.~\ref{fig:AR_2_real_space_fields}.   Note however that in the simulations of Brueur and Hansen~\cite{Breuer:EPL2009} for $\Gamma=2$ and $Pr\rightarrow \infty$ fluid, the most dominant modes were $ (1,1)$ and $ (2,1)$.     
 Thus, a variation of the Prandtl number for the same box can change the flow pattern.
 For our simulations the average intervals between consecutive reversals for $\Gamma = 1$ and $\Gamma =2$ are approximately 0.6 and 0.05 thermal diffusive time units respectively.   The duration of the reversals is approximately 0.03 time units for both the cases.

The flow reversals are not observed for $\Gamma=1$ and $Ra = 10^8,10^9$, and for $\Gamma=2$ and $Ra = 2 \times 10^7, 10^8$ till several thermal diffusive time units (maximum 10), though they may occur after some more time,  a result consistent with that of Sugiyama {\em et al.}~\cite{Sugiyama:PRL2010}.  As the Rayleigh number is increased, the flow reversals become more  difficult due to suppression of the corner rolls ($(2,2)$ mode) by the dominant roll structure, which is quantified by the  ratio of the time-averaged Fourier amplitudes of these modes.    For our simulations on $\Gamma=2$, the ratio  $\langle |\hat{v}_{2,2}|  \rangle_t / \langle |\hat{v}_{2,1}| \rangle_t$  ranges from 0.12 for $Ra=10^7$ (reversing) to 0.04 for $Ra = 10^8$ (non-reversing), which is consistent with the convective flow profiles exhibited in  Fig.~\ref{fig:AR_2_real_space_fields}(a,b)  for these two cases.   For $\Gamma=1$, the corresponding ratio ranges from 0.44 for $Ra=2\times10^7$ to 0.12 for $Ra=10^9$. 

We deduce interesting features on the generation and symmetry of the Fourier modes using the structure of the two-dimensional RBC equations in the Fourier space~\cite{Verma:PRAMANA2006}:  
\begin{eqnarray}
    \frac{\partial \hat{u}_i(\mathbf{k}) }{\partial t} & = & 
    -i k_j \sum_{\bf k=p+q}   \hat{u}_j ({\bf p}) \hat{u}_i({\bf q}) - i k_i P({\bf k}) \nonumber \\
    & &  + Ra Pr \hat{\theta}(\mathbf{k}) \delta_{i,2} 
    	  - Pr k^2  \hat{u}_i(\mathbf{k})  \label{eqn:vel_k_eqn}\\ 
    \frac{\partial \hat{\theta}(\mathbf{k})}{\partial t} & = &
     - i k_j \sum_{\bf k=p+q}   \hat{u}_j ({\bf p}) \hat{\theta}({\bf q}) +  \hat{u}_2(\mathbf{k}) - k^2
    \hat{\theta}(\mathbf{k}) \label{eqn:temp_k_eqn}
\end{eqnarray}
where $\theta$ is the perturbation of the temperature about the conduction state.  For a two-dimensional box, the Fourier modes (coefficients of the basis functions) of the fields are of the type $E$ = (even, even), $O$ = (odd, odd), $M_{eo}$ = (even, odd), and  $M_{oe}$ = (odd, even), where we refer to $E$, $O$ and $M$ as even, odd and mixed modes respectively.  The nonlinear interactions generate new modes with ${\bf k = p + q}$.  For example, modes $(m_1, n_1)$ and $(m_2, n_2)$ generate modes $(|m_1\pm m_2|, |n_1 \pm n_2|)$  (note that $\sin(m x)$ is a superposition of both $k_x = \pm m$).  The nonlinear interactions satisfy the following properties: $O \times O = E$; $O \times E = O$; $E \times E = E$; $E \times M_{eo, oe} = M_{eo, oe}$; $O \times M_{eo ,oe} = M_{oe, eo}$; 
$M_{eo,oe} \times M_{eo,oe} = E$; and $M_{eo,oe} \times M_{oe,eo} = O$.  Here $O \times O = E$ means that two $O$-modes interact to yield an $E$-mode.   As a result of these properties, the two symmetry operations that keep Eqs.~(\ref{eqn:vel_k_eqn}, \ref{eqn:temp_k_eqn}) invariant are: 
\begin{enumerate}
\item $ (E \rightarrow E,  O \rightarrow -O, M_{eo} \rightarrow M_{eo}, 
	M_{oe} \rightarrow -M_{oe})$
 \item $ (E \rightarrow E,  O \rightarrow -O, M_{eo} \rightarrow -M_{eo}, 
	M_{oe} \rightarrow M_{oe})$ 
 \end{enumerate}
 That is, for case (i), if $\{ E, O, M_{eo}, M_{oe} \}$ is a solution of Eqs.~(\ref{eqn:vel_k_eqn}, \ref{eqn:temp_k_eqn}), then $\{ E, -O, M_{eo}, -M_{oe} \}$ is also a solution of these equations.   The system explores  the solution space allowed by these symmetry properties.   Note that the above properties are universal, that is, they are independent of the box geometry, Prandtl number, etc.

To relate our simulations with the above mentioned symmetry, we observe that the $(1,1)$ (of $O$-type)  and $(2,2)$ (of $E$-type) modes are the primary modes for the $\Gamma=1$ system.
Our simulations show that the mode $(1,1)$ and other $O$-type modes switch sign after a reversal, while the sign of $(2,2)$ and other $E$-type modes remains unchanged.  The $M$-type modes have very small energy, hence both the symmetries reduce to one, which is the symmetry of the $\Gamma=1$ system.  For the $\Gamma=2$ box, the primary modes are $(2,1)$ (of $M_{eo}$-type) and $(2,2)$ (of $E$-type).   These primary modes and  subsequently generated modes satisfy the second symmetry mentioned above since the $M_{eo}$-type modes flip, while the $E$-type modes do not.  By analogy, we expect the solution of a $\Gamma=1/2$ box (with finite $Pr$) to satisfy the first symmetry, while the modes of Breuer and Hansen~\cite{Breuer:EPL2009} satisfy the second symmetry (with $(1,1)$ and $(2,1)$ as primary modes).   These symmetry arguments are general, and thus useful for understanding the dynamics of the Fourier modes.  

We also point out that in magnetohydrodynamics (MHD), $\{ {\bf u} \rightarrow {\bf u}, {\bf b} \rightarrow -{\bf b} \}$ is a symmetry of the MHD equations, so all the Fourier modes of the magnetic field ${\bf b}$ change sign after the reversal.   The dynamical equations of RBC do not have such global symmetry.   This is one of the critical differences between flow reversals of RBC and magnetic field reversals of dynamo.

To conclude,  our numerical simulations and symmetry arguments of RBC demonstrate 
the usefulness of large-scale Fourier modes in describing various features of flow reversals, such as the large-scale diagonal roll and corner rolls of Sugiyama {\em et al.}~\cite{Sugiyama:PRL2010}, and the cessation-led reversals reported by Brown and Ahlers~\cite{Brown:JFM2006} and Mishra {\em et al.}~\cite{Mishra:JFM2011}.  We also find that the Nusselt number fluctuates wildly during the flow reversals.   We exploit the structures of the RBC equations in the Fourier space to identify its symmetry transformations.  The symmetry arguments and the reversal dynamics described in this paper are quite general, and they would be useful in understanding of reversals in three-dimensional convective flow as well as in dynamo.  

\begin{acknowledgments}
We are grateful to Paul Fischer and other developers of Nek5000 for opensourcing Nek5000  as well for providing valuable assistance during our work.  We thank Stephan Fauve, Supriyo Paul, Pankaj Mishra,  and Sandeep Reddy for very useful discussions. We also thank the Centre for Development of Advanced Computing  (CDAC) for providing us computing time on  PARAM YUVA.  Part of this work was supported by Swarnajayanti fellowship to MKV, and BRNS grant BRNS/PHY/20090310.
\end{acknowledgments}


\end{document}